\documentclass[reprint,prd,amsmath,superscriptaddress,twocolumn,amssymb,aps,bibliography=totoc]{revtex4}[12pt]

\usepackage[nolist, nolinebreak]{acronym}	

\usepackage[usenames, dvipsnames, svgnames]{xcolor}
\usepackage[pdfview = FitH, pdfstartview = FitH, pdfremotestartview = FitH, pdfduplex = DuplexFlipLongEdge]{hyperref}
\hypersetup{
}

\usepackage{amsmath, amssymb, amsfonts}
\usepackage{epstopdf}
\usepackage{epsfig}
\usepackage[loose]{units}
\usepackage{latexsym}
\usepackage{verbatim}
\usepackage{dhucs}
\usepackage{dcolumn}
\usepackage{bm, bbm}	
\usepackage{notes2bib}
\usepackage{appendix}
\usepackage{slashed}
\usepackage{multirow}
\usepackage{color}
\usepackage{mathtools}
\usepackage{graphicx}
\usepackage{braket}
\usepackage{setspace}
\usepackage{graphicx}
\usepackage{hyperref}
\usepackage{starfont}
\usepackage{bbold}
\usepackage[shortlabels]{enumitem}
\usepackage{framed}
\usepackage{empheq}
\usepackage{amsmath}
\usepackage{tikz-feynman}
\usepackage{simpler-wick}
\usepackage{wasysym}
\usepackage{mhchem}
\usepackage{soul}
 
\def\bea{\begin{eqnarray}}
\def\eea{\end{eqnarray}}

\begin{document}

\title{Searching for a solar relaxion/scalar with XENON1T and LUX}

\author{Ranny Budnik}
\affiliation{Department of Particle Physics and Astrophysics, Weizmann Institute of Science, Rehovot 7610001, Israel}
\author{Oz Davidi}
\affiliation{Department of Particle Physics and Astrophysics, Weizmann Institute of Science, Rehovot 7610001, Israel}
\author{Hyungjin Kim}
\affiliation{Department of Particle Physics and Astrophysics, Weizmann Institute of Science, Rehovot 7610001, Israel}
\author{Gilad Perez}
\affiliation{Department of Particle Physics and Astrophysics, Weizmann Institute of Science, Rehovot 7610001, Israel}
\author{Nadav Priel}
\affiliation{Department of Particle Physics and Astrophysics, Weizmann Institute of Science, Rehovot 7610001, Israel}
\affiliation{Department of Physics, Stanford University, Stanford, CA 94305}

\begin{abstract}
We consider liquid xenon dark matter detectors for searching a light scalar particle produced in the solar core, specifically one that couples to electrons. Through its interaction with the electrons, the scalar particle can be produced in the Sun, mainly through Bremsstrahlung process, and subsequently it is absorbed by liquid xenon atoms, leaving prompt scintillation light and ionization events. Using the latest experimental results of XENON1T and Large Underground Xenon, we place bounds on the coupling between electrons and a light scalar as \(g_{\phi ee} < 7 \times 10^{-15}\) from S1-only analysis, and as \(g_{\phi ee} < 2 \times 10^{-15}\) from S2-only analysis. These can be interpreted as bounds on the mixing angle with the Higgs, \(\sin \theta < 2 \times 10^{-9} \, \left(7 \times 10^{-10}\right)\), for the case of a relaxion that couples to the electrons via this mixing. The bounds are a factor few weaker than the strongest indirect bound inferred from stellar evolution considerations. 
\end{abstract}
\maketitle

\section{Introduction}
Weakly coupled light states arise in a wide variety of beyond the \ac{SM} scenarios. Possibly the two most common examples for such states are light fermions, with their masses being protected by chiral symmetries, and \acp{ALP}, being pseudo-Nambu-Goldstone bosons, their masses are protected by shift symmetries. The \ac{SM} itself, in fact, contains fermion of the above type, the neutrinos, while the most celebrated \ac{ALP} model is that of the QCD axion. In this work, however, we focus on the case of a weakly coupled scalar particle. This case has received considerably less attention from the community, and rightly so. As is well known, light scalars are subject to additive renormalization from ultraviolet scale, and are thus generically less motivated. There are, however, known quantum field theories that can protect the scalars from radiative contributions, such as models with conformal symmetry, and supersymmetric theories. In addition, there is the possibility that the theory consists of a light \ac{ALP}, where the sector that breaks the shift symmetry also breaks parity. This implies that the \ac{ALP} is no longer an eigenstate of parity (or CP), and can acquire both even and odd couplings. The relaxion field that was proposed to address the hierarchy problem~\cite{Graham:2015cka} is an interesting example of the above kind. The relaxion mass is protected by a shift symmetry that is broken by two sequestered sources, implying that its vacuum expectation value is generic and parity is spontaneously broken~\cite{Flacke:2016szy,Choi:2016luu}. It generically thus gives rise to a mixing between the relaxion and the \ac{SM} Higgs that leads to a variety of experimental signatures~\cite{Flacke:2016szy}. Obviously, such a mixing can also be present in generic models of Higgs-scalar portal (see for instance Ref.~\cite{Beacham:2019nyx} and Refs. therein).

The above provides us with a motivation to search for such a scalar state, a singlet of the \ac{SM} gauge interactions. For scalar masses of a few MeV to the few hundred GeV scale, collider searches, beam dump experiments, and flavor experiments constrain the mixing angle between the Higgs and the scalar state. Below the MeV scale, astrophysical/cosmological observations, and searches for violation of the equivalence principle provide the strongest constraints (see \emph{e.g.} Refs.~\cite{Krnjaic:2015mbs,Choi:2016luu,Flacke:2016szy,Frugiuele:2018coc,Fradette:2018hhl} and references therein). If the scalar particle constitutes \ac{DM} in the present universe, and if its mass is sub-eV scale, a coherent oscillation of scalar \ac{DM} induces an oscillation in fundamental constants, and thus, precision atomic sensors can be used as an alternative probe~\cite{Arvanitaki:2014faa,Graham:2015ifn,Arvanitaki:2017nhi,Safronova:2017xyt,Banerjee:2018xmn,Aharony:2019iad,Antypas:2019qji}. 

In this paper, we investigate the production of a light scalar particle from the Sun. Specifically, we use the data of XENON1T and \ac{LUX} to constrain the scalar parameter space. In principle, there could be various types of relevant interactions with \ac{SM} particles, \emph{e.g.} a coupling to electrons, a coupling to nucleons or a coupling to photons. Couplings to nucleons and photons open different channels for the scalar production in the Sun, such as ion-ion Bremsstrahlung and/or Primakoff production. However, these couplings are irrelevant for the detection in \ac{LXe} detectors. \ac{LXe} detectors are not sensitive to the coupling to photons. Moreover, the produced scalar particle has energies of \(\unit[\mathcal{O}(1)]{keV}\), which is the core temperature of the Sun, and, for this energy scale, nuclear absorption is not possible, while the elastic scattering yields very low recoil energy, below present day detector thresholds. On the other hand, electrons are abundant in the solar core, which can open production channels such as electron-ion Bremsstrahlung or Compton-like scattering. Furthermore, an incoming scalar with a typical solar energy can interact with the \ac{LXe} electrons, leaving  easily detectable signals of prompt scintillation and ionization electrons. Thus, the scalar absorption due to its electron coupling will be our main signal, and for this reason, we only consider the electron coupling in our analysis.

The relevant Lagrangian is thus
\bea
{\cal L} \supset -g_{\phi ee} \phi \bar{e} e\,.
\label{int}
\eea
We restrict our attention to sub-keV mass range of these particles, \(m \lesssim \unit[1]{keV}\), such that it can be copiously produced in the Sun. 
Once the light particle is produced, it easily escapes the Sun, as it only weakly couples to the \ac{SM} particles, and eventually reaches the \ac{LXe} detectors. The light scalar particle can be absorbed by xenon atoms, in a process analogous to the axio-electric effect~\cite{Dimopoulos:1985tm,Avignone:1986vm}, which is observed as an electronic-recoil signal of the full energy of the scalar in the detector. Our goal in this paper is to use this signal in \ac{LXe} detectors to probe the light scalar particle of sub-keV mass range.
It is worth noting that using the Sun as a source of weakly interacting light particles is not a new idea. Similar ideas have already been proposed in the past to probe  other weakly coupled light states, such as axions and \acp{ALP}~\cite{Armengaud:2013rta,Aprile:2014eoa,Akerib:2017uem,Ahmed:2009ht,Yoon:2016ogs,Abe:2012ut}, and dark photons~\cite{An:2013yua, Bloch:2016sjj,Hochberg:2016sqx}, with \ac{DM} direct detection experiments.

The paper is organized as follows. In Sec.~\ref{sec:production}, we discuss the production of light scalar particles from the Sun. We determine the relevant processes for the production, and compute the flux resulting from these processes. In Sec.~\ref{sec:analysis}, we present detailed analysis with data taken from XENON1T and \ac{LUX}. Then, we discuss our results in comparison with other existing constraints on the same coupling in Sec.~\ref{sec:result_discussion}. We also discuss the implications of our result in the context of several new physics scenarios in the same section.

\section{Solar Production}\label{sec:production}
We investigate the production of a light scalar particle in the Sun. The total differential production rate is the sum of differential production rates by each process,
\bea
\Gamma^{\rm prod} = \Gamma^{\rm bb} + \Gamma^{\rm bf} + \Gamma^{\rm ff} + \Gamma^{\rm ee} + \Gamma^{\rm C}\,,
\label{eq:Production Rate}
\eea
where bb is the production rate from transitions of bounded electrons, bf is from recombination of free electrons with ions, ff is Bremsstrahlung emission due to scatterings of electrons on ions, ee is Bremsstrahlung emission due to scatterings of two electrons, and C is Compton-like scattering \(\gamma + e \to \phi + e\). To find the total differential flux, we integrate the differential production rate over the solar volume. The total differential flux is given as
\begin{equation}
\frac{d\Phi}{d\omega} = \frac{\omega k}{8\pi^3 R^{2}} \intop_{\odot}  dV \, \Gamma^{\rm prod} \!\left(\omega\right)\,,
\label{eq:Differential Particle Flux}
\end{equation}
where \(R = \unit[1]{AU}\) is the distance between the Sun and the Earth.

The Bremsstrahlung process is the dominant production mechanism  for the relevant energy scale. The electron-electron Bremsstrahlung is a relativistic process, and hence is suppressed compared to the electron-ion Bremsstrahlung, \(\Gamma^{\rm ff}\), thus we take into account only the latter. To find the Bremsstrahlung rate for the scalar particle, we may compute the matrix element and the production rate directly, but a more easy and physically intuitive way to obtain the rate is to first observe a relation between the matrix elements for emitting one photon and one scalar particle from an electron. The ratio between the matrix elements of the two processes is~\cite{Avignone:1986vm}
\bea
\frac{\left|\mathcal{M}\left(e \rightarrow e + \phi\right)\right|^{2}}{\left|\mathcal{M}\left(e \rightarrow e + \gamma\right)\right|^{2}} 
\simeq \frac{g_{\phi ee}^2  v_\phi^2}{4\pi \alpha_{\textrm{em}}}\,,
\label{eq:Scalar Electric Ratio}
\eea
where \(v_{\phi} = k / \omega\) is the velocity of the emitted scalar particle, \(\alpha_{\textrm{em}}\) is the electromagnetic fine structure constant, and the matrix element in the denominator is obtained after averaging over photon polarization states. As was already mentioned by the authors of Ref.~\cite{Avignone:1986vm}, Eq.~\eqref{eq:Scalar Electric Ratio} exhibits an \(m_{e}^{2} / \omega^{2}\) enhancement compared to the analogous ratio for \acp{ALP}. From the above observation, we find the production rate due to electron-ion Bremsstrahlung to be
\bea
\Gamma^{\rm ff}\!\left(\omega\right) = \frac{g_{\phi ee}^2 v_\phi^2 }{4\pi \alpha_{\rm em}}  \Gamma_\gamma^{\rm ff}\!\left(\omega\right)\,,
\label{ff}
\eea
where the Bremsstrahlung rate for photon \(\Gamma_\gamma^{\rm ff}\) is given by~\cite{Redondo:2013lna}
\bea
\Gamma_\gamma^{\rm ff}\!\left(\omega\right) = \alpha_{\rm em}^3 \frac{64\pi^{2}}{3\sqrt{2\pi}}\frac{n_{e} \sum_{j}n_{j} Z_j^2}{\omega^{3}m_{e}^{3/2}\sqrt{T}}e^{-\omega/T} F\!\left(\omega /T\right)\,.
\label{brem}
\eea
Here, \(n_e\) is the electron density, \(n_j\) is the density of atoms with atomic number \(Z_j\), \(T\) is the temperature of the Sun, and the function \(F\), corresponding to the Gaunt factor in the Born approximation up to some constant factor, is defined as
\bea
F\!\left(a\right) = \intop^{\infty}_{0} dx \, x \, e^{-x^{2}}\intop^{\sqrt{x^{2} + a} + x}_{\sqrt{x^{2} + a} - x} dt \, \frac{t^{3}}{\left(t^{2} + y^{2}\right)^{2}}\,,
\eea
where \(y = k_s / \sqrt{2m_e T}\), and 
\begin{equation}
k_{s} = \sqrt{\frac{4\pi\alpha_{\textrm{em}}}{T} \sum_{j} n_{j}} Z_{j}^{2}
\end{equation}
is the Debye screening scale. Although Eq.~\eqref{brem} should be, in principle, summed over all elements inside the Sun, the dominant contribution arises from electron-proton, and electron-\(\alpha\)-particle scattering.

The production rate for the Compton-like process can be easily computed as well, using the scalar-electric relation, Eq.~\eqref{eq:Scalar Electric Ratio}. 
We find
\bea
\Gamma^{\textrm{C}}\!\left(\omega\right) =  \frac{g_{\phi ee}^2 v_\phi^2}{4\pi\alpha_{\textrm{em}}} \Gamma^{\textrm{C}}_\gamma\!\left(\omega\right)\,,\label{eq:Compton-Like Production Rate}
\eea
where \(\Gamma_{\gamma}^{\textrm{C}}\!\left(\omega\right) = f\!\left(\omega\right) n_{e} \sigma_{\textrm{T}}/2\) is the differential production rate of photons through Compton scattering, \(f\!\left(\omega\right) = 1/\left(e^{\omega/T} - 1\right)\) is the phase space distribution of the photon, and \(\sigma_{\textrm{T}} = \left(8\pi \alpha_{\textrm{em}}^2 / 3m_e^2\right)\) is the Thomson scattering cross-section. We have confirmed the expressions of Eq.~\eqref{ff} and Eq.~\eqref{eq:Compton-Like Production Rate} by performing a direct computation in the non-relativistic limit.

The production rate from atomic transitions is more complicated, as it involves computation of matrix elements of atomic transitions in the thermal bath.
Instead of directly computing the matrix elements, we may use available data for photon absorption rates in the Sun, and infer the production rate of the photon by using detailed balance. Applying the scalar-electric relation, the production rate of the scalar particle can then be obtained.
The photon absorption rate is already computed in the context of radiative transport. The intensity of photons at frequency \(\omega\) evolves according to the equation
\begin{equation}
\frac{dI\!\left(\omega\right)}{ds} = - k\!\left(\omega\right) I\!\left(\omega\right) + j\!\left(\omega\right)\,,
\end{equation}
where \(s\) is a coordinate along the line of sight, \(k\!\left(\omega\right)\) is an absorption coefficient, and \(j\!\left(\omega\right)\) is a source. This absorption coefficient can be written as a function of photon absorption rate as \(k\!\left(\omega\right) = \!\left(\Gamma_\gamma^{\rm ff} + \Gamma_\gamma^{\rm bf} + \Gamma_\gamma^{\rm bb}\right)_{\rm abs} \left( 1 - e^{-\omega/T}\right) + \Gamma^{\rm C}_{\gamma,\rm abs}\). 
Then, the absorption rate can be translated into the production rate by detailed balance, \(\Gamma_{\rm abs} = e^{\omega/T} \Gamma_{\rm prod}\). Using Eq.~\eqref{eq:Scalar Electric Ratio}, we can find the scalar production rate from atomic transition as a function of \(k\!\left(\omega\right)\)
\bea
\Gamma^{\rm bb} + \Gamma^{\rm bf} + \Gamma^{\rm ff} = \frac{g_{\phi ee}^2 v_\phi^2}{4\pi\alpha_{\textrm{em}}} f(\omega) \left[ k(\omega)  - e^{\omega/T} \Gamma_\gamma^{\textrm{C}} \right]\,.
\label{op_rel}
\eea
The photon absorption coefficient, \(k\!\left(\omega\right)\), can be extracted from simulated data of photon opacity inside the Sun~\cite{Badnell:2004rz,Seaton:2004uz} (see also Ref.~\bibnote{\url{http://cdsweb.u-strasbg.fr/topbase/TheOP.html}}). This method has already been used to compute the axion flux from the solar core~\cite{Redondo:2013wwa}.

\begin{figure}[t]
\centering
\includegraphics[scale=0.4]{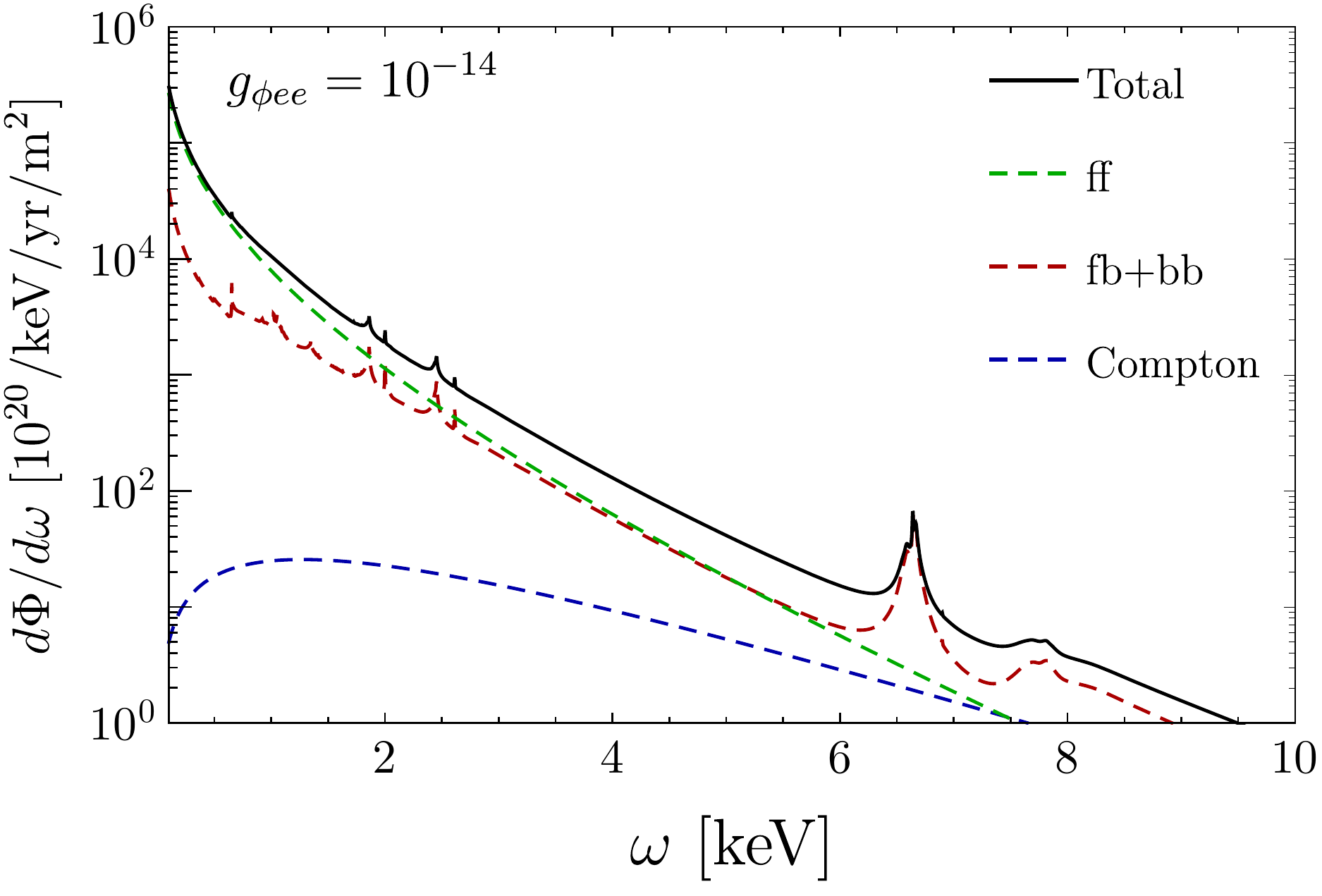}
\caption{%
Scalar flux from the Sun. The coupling to electrons is chosen to be \(g_{\phi ee} = 10^{-14}\), and the scalar is taken to be massless. The green dashed line is Bremsstrahlung flux from electrons interacting with hydrogen and helium ions, the red dashed line is the flux due to recombination and transitions of bounded electrons in heavier elements, the blue dashed line is the flux from Compton-like scattering, and black line is the total scalar flux.
}
\label{fig:flux}
\end{figure}

Using these results, we integrate the differential production rate over the solar volume by using the solar model obtained in Ref.~\cite{Vinyoles:2016djt}. In Fig.~\ref{fig:flux}, we present the light scalar flux. The Bremsstrahlung processes, involving hydrogen and helium, are calculated directly by using Eq.~\eqref{brem}, while atomic transition processes for these atoms are neglected, because all energy levels of these elements lie below our experimental threshold. The flux from interactions between electrons and heavy elements, which include C, N, O, Ne, Mg, Si, S, and Fe, is obtained from the opacity relation, Eq.~\eqref{op_rel}. We use the massless limit for the scalar, and take coupling constant \(g_{\phi ee} = 10^{-14}\) for this plot.

We finally comment on the resonant production of light scalars in the plasma. The scalar field with interaction Eq.~\eqref{int} can mix with the longitudinal excitation of the photon in the plasma~\cite{Hardy:2016kme} (see also Refs.~\cite{An:2013yfc,Redondo:2013lna} for earlier studies on the mixing of dark photon with the longitudinal photon mode in the Sun). In the presence of such mixing, the production rate of the scalar field changes as
\begin{equation}\label{eq:Resonant Enhancement}
\Gamma^{\textrm{prod}}\!\left(\omega\right) = \frac{g_{\phi ee}^{2}v_{\phi}^{2}}{4\pi\alpha_{\textrm{em}}}\frac{f\!\left(\omega\right)\Gamma_{L}\!\left(\omega\right)}{\left(1-\frac{\omega_{\textrm{p}}^{2}}{\omega^{2}}\right)^{2} + \left(\frac{\Gamma_{L}}{\omega}\right)^{2}}\,,
\end{equation}
where \(\omega_{\textrm{p}}^2 = 4 \pi \alpha_{\textrm{em}} n_e / m_e\) is the plasma frequency, and \(\Gamma_{L} = \Gamma_L^{\rm abs} - \Gamma_L^{\rm prod}\) is the damping rate of the longitudinal excitation of the photon. The term on the numerator is the production rate of the longitudinal mode, \(\Gamma_{L}^{\rm prod} = f\!\left(\omega\right) \Gamma_L\), which can be seen by using detailed balance. The resonance takes place when the frequency of the scalar particle is close to the plasma frequency, \(\omega \approx \omega_{\textrm{p}}\), and numerically, the flux at this frequency could be enhanced roughly by two or three orders of magnitude compared to the non-resonant case.
However, the plasma frequency inside the Sun ranges as \(\unit[1]{eV} \lesssim \omega_{\textrm{p}} \lesssim \unit[300]{eV}\), and thus, the frequency of resonantly produced scalar particles is also limited from above by \(\unit[300]{eV}\). 
This energy scale is too small for the analysis of prompt scintillation signals, but is relevant for analysis of ionization signals in \ac{LXe} detectors, as we will discuss below.

\section{Scalar absorption in detectors}\label{sec:analysis}
Once the light states are produced in the Sun, they pass through the detectors, and are expected to ionize atoms due to the same interaction, Eq.~\eqref{int}. This effect is similar to the photoelectric effect. The expected event rate is obtained by the convolution of the solar flux, Eq.~\eqref{eq:Differential Particle Flux}, with the absorption cross-section. Instead of directly computing the absorption cross-section, we use again the relation between matrix elements, Eq.~\eqref{eq:Scalar Electric Ratio}, and estimate the absorption cross-section as a function of the photoelectric cross-section,
\begin{equation}
\frac{\sigma\!\left(\omega\right) v_{\phi}}{\sigma_{\textrm{pe}}\!\left(\omega\right) c} = \frac{g_{\phi ee}^{2} v_{\phi}^{2}}{4\pi\alpha_{\rm em}}\,,
\label{eq:Scalar Electric Effect}
\end{equation}
where \(\sigma_{\textrm{pe}}\) is the photoelectric cross-section for \ac{LXe}. We ignore mixing of scalar particle with longitudinal excitation in the detector, since we are interested in scalar particles with \(\omega \simeq \unit[{\cal O}\!\left(1\right)]{keV}\), and for this frequency range, the mixing is barely important. For the photoelectric cross-section, we take tabulated data of \(\sigma_{\rm pe}\) from Refs.~\cite{Veigele:1973tza,XCOM}.

Having determined the flux and absorption cross-section, we use three sets of data, obtained by the XENON1T and \ac{LUX} collaborations, to set limits on the coupling of scalar particles to electrons. We first discuss the prompt scintillation signals (denoted as S1).
We use \ac{LUX} data, collected in 2013 with an exposure of \(\unit[95]{days}\) and \(\unit[118]{kg}\) of fiducial mass, which is the same data set used by \ac{LUX} collaboration to search for solar axions~\cite{Akerib:2017uem}. In addition, we use the XENON1T data set with an exposure of \(\unit[1]{ton\times yr}\), which has been used to probe \ac{WIMP} \ac{DM}~\cite{Aprile:2018dbl}.

The detector response is taken into account in the following way. First, the deposited energy is translated into expected S1 signal using \({\rm S1} = g_1 \omega L_y\). The light yield, \(L_y\), is the expected number of scintillation photons for a given energy deposition, and is taken from Ref.~\cite{Akerib:2016qlr} both for the \ac{LUX} and XENON1T analyses. The detection efficiency for photons, \(g_{1}\), is taken from Ref.~\cite{Akerib:2017uem} for the \ac{LUX} analysis, and from Ref.~\cite{Aprile:2019dme} for the XENON1T analysis.
Then, the observed S1 signal in the detector is obtained by Poisson-smearing and binning the expected signal. 

For this analysis, we place a threshold of \unit[3]{\ac{PE}} in S1, and set \(L_{y}\) to zero below the lowest measured data point of \(\unit[1.3]{keV}\). This procedure is established in order to avoid oversensitive response to the steeply raising solar flux below \(\unit[1]{keV}\), and can be relaxed in a future more elaborated analysis.

Several systematic studies have been performed to quantify the effect of below-threshold energy deposition. In particular, the light yield was extrapolated down to zero energy, the Poisson smearing was replaced by a Gaussian one, following the procedure described in Ref.~\cite{Akerib:2016qlr}, and the results were compared to no smearing with a hard cut at \(\unit[1.5]{keV}\) (which is equivalent to \unit[3]{\ac{PE}}). For all of these studies, the resulting limits have changed by less than 5\%.

A profile likelihood procedure~\cite{Cowan:2010js} has been used to calculate the upper bounds on \(g_{\phi ee}\) at 90\% \ac{CL}. We use a binned likelihood function, which is a product of the Poisson probability of each bin (\emph{c.f.} Ref.~\cite{Patrignani:2016xqp}).

The expected number of signal events in each energy bin is estimated as described in the preceding section, and the expected number of background events in each bin is taken from Ref.~\cite{Akerib:2017uem} for \ac{LUX}, and from Ref.~\cite{Aprile:2018dbl} for XENON1T. For \ac{LUX}, we use 11 bins in the range of \unit[3-60]{\ac{PE}}, and for XENON1T we use a single bin in the range of \unit[3-10]{\ac{PE}}. As the XENON1T data is not binned, we estimate the background by assuming a flat background in S1 space. This assumption results in 66 expected background events in the search region. The number of data points in the search region was estimated to be 70.

\begin{figure}[t]
	\centering
	\includegraphics[scale=0.42]{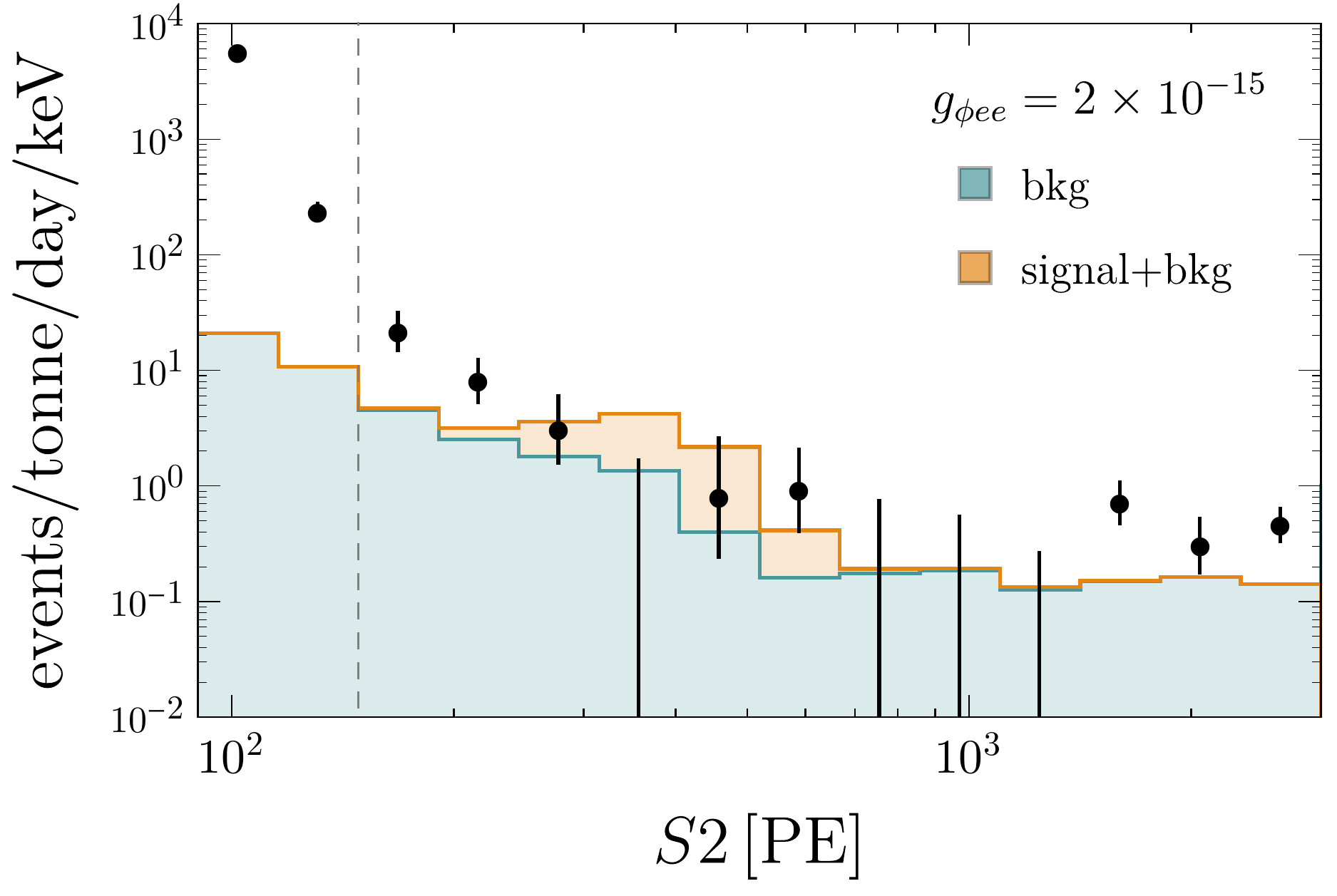}
	\caption{
		The black dots are the data with error bars showing statistical uncertainties (\(1 \sigma\) Poisson), the blue shaded histogram is the partial background model from XENON1T experiment, and the vertical dashed line shows the S2 threshold~\cite{Aprile:2019xxb}.
		The orange shaded histogram, stacked on the background, is the signal model for a massless scalar, and \(g_{\phi ee} = 2 \times 10^{-15}\).
		There is a peak around S2 = \unit[300--400]{\acs*{PE}}, which corresponds to the resonant peak of the scalar spectrum at \(\omega \sim \unit[0.2]{keV}\).
	}
	\label{fig:signal_model}
\end{figure}

We have also performed an S2-only analysis.
The scalar flux produced in the Sun is enhanced at low energies by a resonant production, as mentioned above. Therefore, a search with lower threshold will yield much higher sensitivities, given that the background can be modeled, and that its rate is not exponentially raising. This technique has been used before in XENON100~\cite{Aprile:2016wwo}, and recently in XENON1T~\cite{Aprile:2019xxb}, where the threshold was lowered by an order of magnitude with respect to S1-only analysis. We take the XENON1T data set~\cite{Aprile:2019xxb} with an exposure of \(\unit[22]{tonne\textrm{--}days}\) for the S2-only analysis, and use the detector response model presented in Ref.~\cite{Aprile:2019dme} to translate the deposited energy into S2 signals. As in Ref.~\cite{Aprile:2019xxb}, we assume that the electronic-recoil events below \(\unit[186]{eV}\) are undetectable for a conservative estimate. Then, we perform the same likelihood analysis to obtain an upper limit on \(g_{\phi ee}\). The signal model, as well as data points and partial background model from XENON1T, are presented in Fig.~\ref{fig:signal_model}.

\section{Results and Discussions}\label{sec:result_discussion}
\begin{figure}
	\includegraphics[width=1\linewidth]{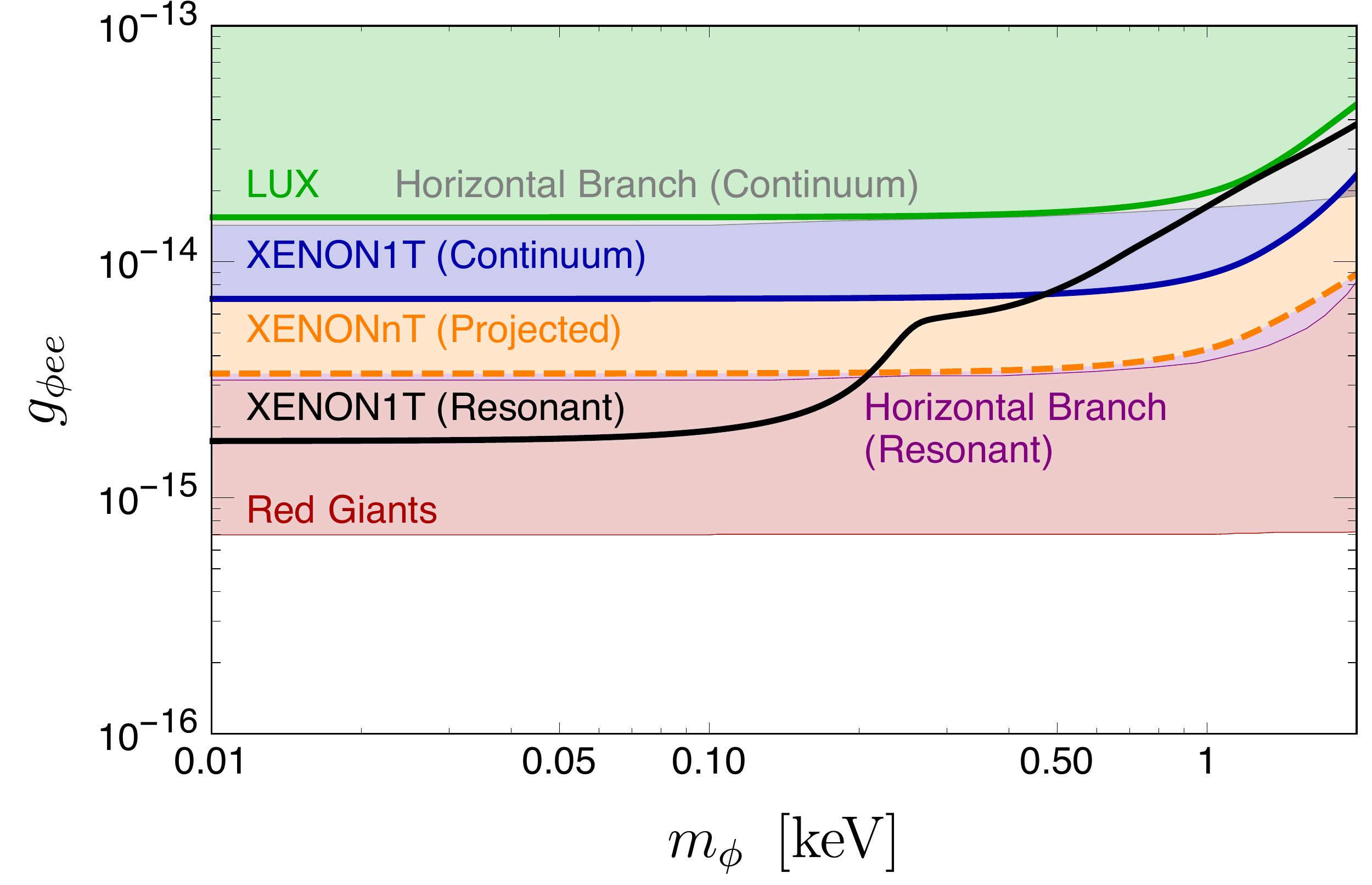}
	\caption{%
		Constraints on the coupling \(g_{\phi ee}\). 
		The result of this work includes the green thick solid line (\acs*{LUX}), the blue thick solid line (XENON1T), the orange thick dashed line (projected sensitivity of XENONnT experiment with \(\unit[20]{ton \times yr}\) exposure), and the black thick solid line (XENON1T S2-only analysis, capturing the resonant production). 
		The thin red, purple, and gray lines represent stellar cooling bounds from red giants, horizontal branch stars with and without in-medium mixing effect~\cite{Hardy:2016kme}, respectively.
	}
	\label{fig:ge_limit}
\end{figure}

We summarize our results for the S1-only analysis in Fig.~\ref{fig:ge_limit}. 
In the limit where the scalar mass can be ignored, the bound on \(g_{\phi ee}\) becomes
\begin{align}
g_{\phi ee} & < 2 \times 10^{-14} \qquad \textrm{(\ac{LUX})} \\
g_{\phi ee} & < 7 \times 10^{-15} \qquad \textrm{(XENON1T)}
\end{align}
at 90\% \ac{CL}. In addition to these experiments, we also show the projected sensitivity of XENONnT. Since the total event rate at the detector is proportional to \(g_{\phi ee}^{4}\), the expected improvement on the electron coupling only scales as a fourth root of the exposure.
The projected limit of XENONnT with a total exposure of \(\unit[20]{ton\times yr}\), and an assumed flat background rate of \(\unit[1000]{events/ton/year}\)~\cite{Aprile:2015uzo} yields
\begin{equation}
\ g_{\phi ee} < 3 \times 10^{-15} \qquad \textrm{(XENONnT, projected)}\,.
\end{equation}
Again, this bound is valid for \(m_\phi < \unit[1]{keV}\). 
In the same figure, we also present the constraints from the other searches on \(g_{\phi ee}\). 
As we consider new light particles below keV, the most stringent constraints come from astrophysical observations. 
Since the electron coupling to a light scalar state opens additional channels for these astrophysical objects to lose their energy, it can lead to anomalous cooling of stellar objects, such as red giants, and stars on the horizontal branch.
Non-observation of such anomalous cooling of stellar objects could place a constraint on a possible electron coupling to any new light states~\cite{Raffelt:1996wa}.
Compared to the results of earlier studies on scalar-electron coupling, \(g_{\phi ee} < 11 \times 10^{-15}\)~\cite{Grifols:1988fv,Raffelt:1996wa}, our result is a factor few better. However, a recent study has improved the stellar cooling bound on the same coupling to \(g_{\phi ee} < 7 \times 10^{-16}\), when accounting for in-medium mixing effect of the scalar field~\cite{Hardy:2016kme}. 
Compared to the latest stellar cooling bound, our current S1-only limit, obtained from XENON1T data, is about an order of magnitude weaker. 

We also present the result of S2-only analysis. From the recent data of XENON1T experiment~\cite{Aprile:2019xxb}, we find
\bea
g_{\phi ee} < 2 \times 10^{-15} \label{eq:S2-Only Bound}
\eea
at 90\% \ac{CL}. This bound is valid for \(m_\phi \lesssim \unit[0.1]{keV}\), since the constraining power mainly comes from a resonant peak in the signal. The sensitivity of the XENON1T S2-only analysis is a factor three weaker than the strongest stellar cooling bound. Upper limits on \(g_{\phi ee}\) for scalar masses larger than \(\unit[0.01]{keV}\) are presented in Fig.~\ref{fig:ge_limit}

Our result can be interpreted in the context of several theoretically well-motivated new physics scenarios. 
Examples include a cosmological relaxion, and Higgs portal singlet scalar field. 
In both scenarios, a scalar state naturally mixes with the \ac{SM} Higgs boson, and thus becomes coupled to the \ac{SM} particles with coupling constant given by
\begin{equation}
g_{\phi\psi\psi} = g_{\psi} \sin\theta\,,
\end{equation}
where \(g_{\psi}\) is the standard Higgs coupling to SM fermions, and \(\sin\theta\) is the scalar-Higgs mixing angle. For electron coupling, \(g_{e} = \kappa_{e} m_{e} / \upsilon = 3 \times 10^{-6} \left(\kappa_{e} / 1\right)\), where \(\upsilon = \unit[174]{GeV}\) is the electroweak scale, and in the \ac{SM}, \(\kappa_{e}=1\). Direct searches for Higgs decay and resonant production allow for a \(600\) times stronger interaction strength with electrons than the predicted \ac{SM} value~\cite{Altmannshofer:2015qra}, \emph{i.e.} \(\kappa_{e} = 600\). Our bound, Eq.~\eqref{eq:S2-Only Bound}, can be thus interpreted as a bound on the mixing angle
\begin{equation}
\sin\theta < \begin{cases}
7 \times 10^{-10}  & \kappa_{e} = 1   \\
1 \times 10^{-12} & \kappa_{e} = 600
\end{cases}\,.
\end{equation}
This applies to any scalar that mixes with the Higgs.
For Higgs portal models (see \emph{e.g.} Ref.~\cite{Piazza:2010ye}), the scalar mass receives a radiative correction from its mixing with Higgs, \(\Delta m_\phi^2 \simeq \upsilon^2 \sin^2\theta / 16\pi^2\), and thus, a natural model requires
\begin{equation}
	\sin\theta \lesssim \frac{4\pi m_{\phi}}{\upsilon} = 7 \times 10^{-9} \cdot \left(\frac{m_{\phi}}{\unit[100]{eV}}\right)\,,
\end{equation}
hence our bound probes natural models. For a generic scalar particle which couples to the electron, the scalar mass receives a quadratically divergent radiative correction from an electron loop, and the cutoff of the theory is bounded by the same naturalness argument as
\begin{align}
\Lambda_{\textrm{NP}} & \lesssim \frac{4 \pi m_{\phi}}{g_{\phi ee}} \lesssim \unit[6 \times 10^{5}]{TeV} \cdot \left(\frac{m_{\phi}}{\unit[100]{eV}}\right)\,.
\end{align}

In this paper, we have considered solar production of light scalar particles, and used \ac{LXe} detectors, XENON1T and \ac{LUX}, to probe its coupling to electrons.
We have only considered a subset of \ac{DM} direct detectors, which has been mainly focusing on detection of \ac{WIMP} \ac{DM}.
Another interesting direction is to use newly proposed ideas on \ac{DM} direct detection, making use of inter-atomic interactions to lower the threshold (see \emph{e.g.} Refs.~\cite{Battaglieri:2017aum,Lin:2019uvt} and references therein).
Such experiments provide an alternative probe of scalar-electron coupling, as the flux of scalar field from the Sun could be resonantly enhanced.

\medskip

\section{Acknowledgments}
The work of RB is supported by grants from ISF (1295/18), and Pazy-VATAT.
RB is the incumbent of the Arye and Ido Dissentshik Career Development Chair.
The work of GP is supported by grants from the BSF, ERC, ISF, Minerva Foundation, and the Segre Research Award.
NP is partially supported by the Koret Foundation.

\begin{acronym}
	\acro{SM}{Standard Model}
	\acroindefinite{SM}{an}{a}
	\acro{ALP}{axion-like particle}
	\acroindefinite{ALP}{an}{an}
	\acro{WIMP}{weakly interacting massive particle}
	\acro{DM}{dark matter}
	\acro{LXe}{liquid xenon}
	\acroindefinite{LXe}{an}{a}
	\acro{LUX}{Large Underground Xenon}
	\acro{PE}{photoelectrons}
	\acro{CL}{confidence level}
\end{acronym}

\bibliography{solar_bounds}
\bibliographystyle{h-physrev5}
\end{document}